\documentclass[aps,pra,floatfix,twocolumn,superscriptaddress,showpacs,10pt]{revtex4-1}
\usepackage{amssymb, amsmath,color,mciteplus,graphicx,subfigure}
\usepackage{calc,epsfig,epstopdf,color,mciteplus,bm,mathrsfs}

\usepackage{times}

\begin{document}

\title{Dynamically Corrected Nonadiabatic Holonomic Quantum Gates}

\author{Sai Li}

\affiliation{Guangdong Provincial Key Laboratory of Quantum Engineering and Quantum Materials,
and School of Physics\\ and Telecommunication Engineering, South China Normal University, Guangzhou 510006, China}

\author{Zheng-Yuan Xue}\email{zyxue83@163.com}
\affiliation{Guangdong Provincial Key Laboratory of Quantum Engineering and Quantum Materials, 
and School of Physics\\ and Telecommunication Engineering, South China Normal University, Guangzhou 510006, China}
\affiliation{Guangdong-Hong Kong Joint Laboratory of Quantum Matter, and Frontier Research Institute for Physics,\\ South China Normal University, Guangzhou 510006, China}
\date{\today}

\begin{abstract}
The key for realizing fault-tolerant quantum computation lies in maintaining the coherence of all  qubits so that high-fidelity and robust  quantum manipulations on them can be achieved.  One of the promising approaches is to use geometric phases in the construction of universal  quantum gates,  due to their intrinsic robustness against certain types of local noises. However, due to limitations in previous implementations, the noise-resilience feature of nonadiabatic holonomic quantum computation (NHQC) still needs to be improved. Here, combining with the dynamical correction technique, we propose a general protocol of universal NHQC with simplified control, which can greatly suppress the effect of the accompanied \emph{X} errors, retaining the main merit of geometric quantum operations. Numerical simulation shows that the performance of our gate can be much better than previous protocols. Remarkably, when incorporating  a decoherence-free subspace encoding for the collective dephasing noise, our scheme can also be robust against  the  involved \emph{Z} errors. In addition, we also outline the physical implementation of the protocol that is insensitive to both \emph{X} and \emph{Z} errors. Therefore, our  protocol provides a promising strategy for scalable  fault-tolerant quantum computation.
\end{abstract}

\maketitle

\section{Introduction}

Quantum computation and its physical implementation have  attracted much attention soon after the discovery of Shor's algorithm \cite{shor}, which is an efficient algorithm for factorizing prime numbers, a hard problem for classical computers. However, up to now, physical realization of the large-scale and fault-tolerant quantum computation \cite{QS2019} is still challenging due to the inevitable noises and operational errors when manipulating a qubit. Therefore, considering the fact that coherent time of quantum systems is limited, realizing fast  and robust quantum gates with high-fidelity is the key for realizing large-scale fault-tolerant quantum computation.

Geometric phases \cite{Abelian,non-Abelian,AA} are induced on a quantum system by moving a set of its Hamiltonian parameters to form a cyclic evolution in the parametric space, and determined by the trajectory that the quantum system travels. This global property leads to the intrinsic noise-resilient feature of geometric phases, which finds promising application in  quantum computation, where quantum gates are implemented using geometric phases, i.e., the geometric quantum computation \cite{zanardi,AGQC1,Duan,cenlx}. Recently,  based on fast nonadiabatic non-Abelian geometric phases, quantum computation protocols have been proposed \cite{NJP,TongDM} and expanded  extensively \cite{liubj, Singleloopxu, Singleloop, SingleloopSQ, surface1, surface2, surface3, eric, zhaopz2020, liubj2020}, with elementary experimental demonstrations   \cite{Abdumalikov35, Feng39, Zu41, AC2014,nv2017,nv20172,li2017, xuy37} on various quantum systems. However,  due to the  implementation limitations, the preferred merit of being resilience against operational control error, i.e., the \emph{X} error,  of geometric phases is smeared \cite{three1, three2}.

Meanwhile, to deal with general errors in nonadiabatic holonomic quantum computation (NHQC), various protocols have also been proposed with preliminary experimental demonstrations. Specifically, besides the conventional wisdom by using error-avoiding encoding technique \cite{encode, zhangj2014, liang2014,zhouj, xue1, wangym2016, xue2, xue3, zhaopz2017, xue4, wangym2020}, which is resource-inefficient,  incorporating with other control techniques in NHQC has also being introduced,  such as the relatively experimental-friendly composite scheme \cite{composite, zhu2019} or dynamical decoupling strategy \cite{dd, dd2}, the deliberately optimal pulse control technique \cite{Liu18, yan2019, Li, ai2020, ai2021, GPC}, and shorten the gate time method with prescribed complex pulses \cite{xugf2018, zhang2019, Chentoc3, BNHQC, yuyang}, etc. However, in these previous explorations,  the enhancement of the gate robustness is obtained at the cost of deliberate pulse-shape control technique and/or greatly extended the gate-time. Thus, NHQC protocol with simplified control while retaining the strong gate robustness is highly desired.

Here,  combining with the dynamical correction technique \cite{dc1,dc2,dc3}, we propose a general protocol for universal NHQC without deliberate pulse control  and with experimental accessible techniques, termed as DCNHQC, which can improve gate robustness against the  \emph{X} error {from the second order in the conventional NHQC case to the fourth order}, and thus retaining the main merit of geometric quantum gates. Our   gate-time is the same as that of the two-loop composite NHQC (CNHQC) scheme  \cite{zhu2019}, which is much shorter than the NHQC with optimal control (NHQCOC) scheme \cite{Li, ai2020}. However, numerical simulation shows that the gate robustness in our scheme can be much better than both the conventional single-loop NHQC \cite{SingleloopSQ} and the CNHQC. Due to the longer gate-time, the gate-infidelity of NHQCOC  will be much higher than the current scheme, although they share similar performance of gate robustness.  Meanwhile, our scheme can be incorporated with a  decoherence-free subspace (DFS) encoding  for the collective dephasing noise  \cite{dfs1,dfs2,dfs3}, which will be robust against   the collective \emph{Z} error. In this way, the present protocol can be immune to both \emph{X} and \emph{Z} errors. Note that the origin of the gate-robustness enhancement here is different from Ref. \cite{shen}. The enhancement there is due to the much shorter gate-time, and thus the decoherence and errors can only affect the quantum dynamics in a shorter time span.  Here, the robustness of the current scheme is stronger than  Ref. \cite{shen}, but the gate-fidelity here is lower, due to the longer gate-time. Finally, we outline the physical implementation of the protocol that is insensitive to both \emph{X} and \emph{Z} errors, with conventional exchange interactions among physical qubits,  where the unwanted leakage errors can  be effectively eliminated. Therefore, our protocol provides a promising strategy towards large-scale fault-tolerant quantum computation.

\section{Conventional holonomic gates}
To begin with, we first summarize the realization of the conventional single-loop NHQC  \cite{Singleloopxu, Singleloop, SingleloopSQ}, which is an efficient extension of the original NHQC proposal \cite{NJP,TongDM}. We consider a three-level quantum system that is resonantly driven by two different external pulses with amplitude envelopes $\Omega_0(t)$ and $\Omega_1(t)$ and initial phases $\phi_0$ and $\phi_1$, respectively. In the interaction picture, setting $\hbar = 1$ hereafter, the interaction Hamiltonian of the driving quantum system is
\begin{eqnarray}\label{HI}
  \mathcal{H}(t) &=& \left[\Omega_0(t)e^{-i\phi_0}|0\rangle + \Omega_1(t)e^{-i\phi_1}|1\rangle \right]\langle a| + \textrm{H.c.}\notag\\
   &=& \Omega(t) e^{-i\phi_0}|b\rangle\langle a|+ \textrm{H.c.},
\end{eqnarray}
where $\{|0\rangle, |1\rangle\}$ are our chosen computation basis, i.e., the two qubit states, and $|a\rangle$ is an auxiliary state; $\Omega(t) = \sqrt{\Omega^2_0(t)+\Omega^2_1(t)}$, $ |b\rangle = \sin(\theta/2)|0\rangle - \cos(\theta/2)e^{i\phi}|1\rangle$ with $\phi = \phi_0-\phi_1+\pi$, and $\tan(\theta/2)=\Omega_0(t)/\Omega_1(t)$ with $\theta$ being a constant angle, $(\theta, \phi)$  determine the rotation axis of the implemented gates, as illustrated latter. This interacting Hamiltonian has a dark state of $|d\rangle=-\cos(\theta/2)e^{-i\phi}|0\rangle-\sin(\theta/2)|1\rangle$, which is decoupled from the other eigenstates during the quantum dynamics.

\begin{figure}[tbp]
	\begin{center}
		\includegraphics[width=0.8\columnwidth]{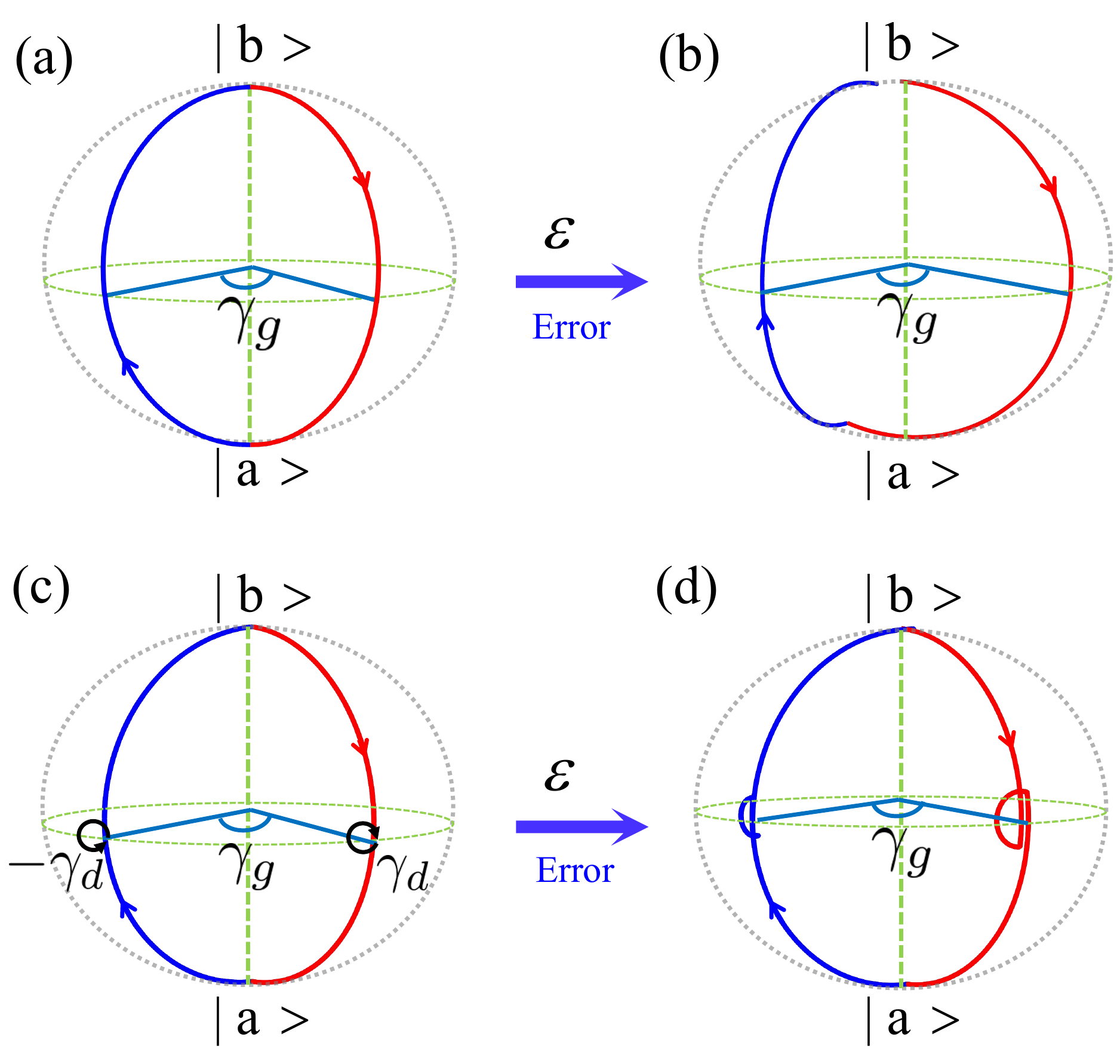}
\caption{Schematic illustration of the evolution paths for holonomic quantum gates in Eq. (\ref{UU}) with $\gamma_g = \pi/2$.  The evolution paths of the bright state $|b\rangle$ for conventional NHQC scheme (a) without and (b) with the systematic \emph{X}  error, which will obviously lead to the gate error.  The evolution paths of the bright state $|b\rangle$ in DCNHQC scheme (c)  without and (d) with the \emph{X}  error.   As shown in (d), the \emph{X}  error can be efficiently eliminated  by the  dynamical correction process. } \label{fig1}
	\end{center}
\end{figure}

With the Hamiltonian above,  conventional NHQC can be realized in a single-loop way, and in the dressed basis $\{|b\rangle, |d\rangle\}$, the cyclic evolution path is shown in Fig. \ref{fig1}(a) and with $\gamma_g = \pi/2$. Specifically, for the first half of the evolution, i.e., $t \in [0, T/2]$ with $T$ being the whole gate-time, under the Hamiltonian in Eq. (\ref{HI})   with $\int^{T/2}_0 \Omega(t) dt =\pi/2$, the bright state $|b\rangle$ starts from the northern pole and travel down to the southern pole along the longitude line determined by $\phi_0$, where it is clockwise rotated an angle $\gamma_g$. Then, it goes back to the northern pole, also along a longitude line, determined by $\phi_0+\pi-\gamma_g$ under the Hamiltonian in Eq. (\ref{HI})   with $\int^{T}_{T/2} \Omega(t) dt =\pi/2$.  After this cyclic evolution, with a total $\pi$ pulse,  $|b\rangle$ state acquires a pure geometric phase $\gamma_g$, while  $|d\rangle$ is not changed, and thus the corresponding holonomic quantum gate operation is
\begin{eqnarray}\label{UU1}
U = |d\rangle \langle d| + e^{i\gamma_g}|b\rangle \langle b|.
\end{eqnarray}
Notably, the $|a\rangle$ state is not our computational state and its population keeps unchanged before and after the gate operation, thus we do not include this component in the above equation. Correspondingly, in the computational basis of $\{|0\rangle, |1\rangle\}$, the holonomic  gate in Eq. (\ref{UU1}) will be
\begin{eqnarray}\label{UU}
U = e^{i{\frac{ \gamma_g}{2}}}e^{-i{\frac{ \gamma_g}{2}}\mathbf{n}\cdot \mathbf{\sigma}},
\end{eqnarray}
where  $\textbf{n}=(\sin\theta\cos\phi,-\sin\theta\sin\phi,\cos\theta)$, and $\mathbf{\sigma}=(\sigma_x, \sigma_y, \sigma_z)$  with $\sigma_{x, y, z}$ are Pauli matrices in the  computational basis. This operation denotes a rotation along the $\textbf{n}$ axis by a angle of $\gamma_g$. As $\textbf{n}$ and  $\gamma_g$ can be arbitrary,  universal single-qubit quantum gates are constructed.

However, as shown in Fig. \ref{fig1}(b), this type of implementation is not robust against  the \emph{X} error. In the presence of  the \emph{X } error, i.e., $  \mathcal{H}^\epsilon(t) = (1+\epsilon) \mathcal{H}(t)$ with $\epsilon$ being the deviation fraction of the amplitude of the driving fields, the evolution can be solved by $U^\epsilon(t)=e^{-i\mathcal{H}^\epsilon(t)t}$. Then, the implemented holonomic gate   in the dressed basis $\{|b\rangle, |d\rangle\}$ changes  to
\begin{eqnarray}\label{Ue1}
U^\epsilon = |d\rangle \langle d| +\left( \cos^2\frac{\mu\pi}{2}+  \sin^2\frac{\mu\pi}{2}e^{i\gamma_g}  \right)|b\rangle \langle b|,
\end{eqnarray}
where $\mu = 1+\epsilon$. Thus, when the \emph{X } error is small, i.e., $\epsilon \ll 1$, the gate fidelity in the dressed basis $\{|b\rangle, |d\rangle\}$ will decrease from the perfect unity to
\begin{eqnarray}\label{Ue1}
F &=& \frac{|Tr(U^\dagger U^\epsilon)|}{|Tr(U^\dagger U)|}=\frac{1}{2}\left|1+\cos^2\frac{\mu\pi}{2}e^{-i\gamma_g}+\sin^2\frac{\mu\pi}{2}\right|\notag\\
& \approx& 1-\epsilon^2\pi^2(1-\cos\gamma_g)/8,
\end{eqnarray}
which shows that the holonomic  quantum gates in the conventional single-loop NHQC scheme can only suppress the  \emph{X} error up to the second order, similar to the performance of dynamical quantum gates \cite{three1, three2}. In this way, when the quantum system suffers from  systematic \emph{X } errors, the $|b\rangle$ state can not  exactly go back to the original starting point after the designed cyclic evolution, as shown in Fig. \ref{fig1}(b), and thus leading to infidelity of the implemented quantum gate.

 \begin{figure}[t]
	\begin{center}
		\includegraphics[width=8.5cm]{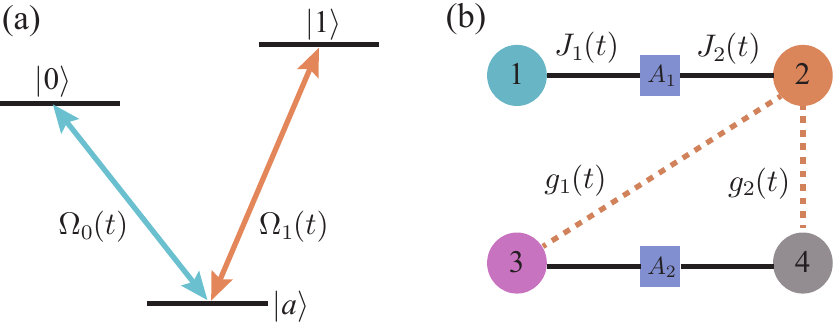}
\caption{ The setup of our proposal. (a)  A $V$ type three-level system with resonantly driving $|0\rangle\leftrightarrow |a\rangle$ and $|1\rangle\leftrightarrow |a\rangle$ transitions. (b) The coupling configuration for a nontrivial two-qubit holonomic quantum gate. The circles with different colors denote the  physical qubits with different frequencies, the squares denote auxiliary physical qubits, and the black solid and orange dashed bonds indicate the interactions are for single and two-qubit gates, respectively. }\label{fig2}
\end{center}
\end{figure}

\section{Dynamically corrected holonomic gates}

We now turn to the implementation of the dynamically corrected holonomic gates. To reduce the influence of the  \emph{X} error in conventional NHQC, under  the guide of the dynamical correction technique \cite{dc1,dc2,dc3}, we insert two dynamical processes to enhance the robustness against  the \emph{X} error in conventional NHQC. Notably, this dynamical correction also works when the two inserted dynamical processes also have \emph{X} errors. Firstly, as shown in Fig. \ref{fig1}(c), we consider the ideal case to explain the evolution process,  where the two dynamical processes are respectively inserted at the middle point in each half of the evolution path. In this way, although there are dynamical phases accumulated during  each dynamical processes,  their summation will be zero, leaving a pure geometric phase $\gamma_g$ on  the dressed state $|b\rangle$, and thus the whole process can still lead  to universal single-qubit holonomic gates in Eq. (\ref{UU}).

Specifically, at the middle of the first half of the evolution path, i.e., at the time $T/4$, a Hamiltonian
\begin{equation}
\mathcal{H}^I_1(t) = \Omega^I_1(t) e^{-i(\phi_0+\pi/2)}|b\rangle \langle a|+ \textrm{H.c.}
\end{equation}
with $\int^{3T/4}_{T/4} \Omega^I_1(t) dt =\pi/2$ is inserted. Due to this inserted pulse, the starting time of the second half of the evolution path changes to $T$, and a second inserted Hamiltonian
\begin{equation}
\mathcal{H}^I_2(t) = \Omega^I_2(t) e^{-i(\phi_0-\gamma_g-\pi/2)} |b\rangle \langle a|+ \textrm{H.c.}
\end{equation}
with $\int^{7T/4}_{5T/4} \Omega^I_2(t) dt =\pi/2$ is turned on at the time $5T/4$. Therefore, in this case, the total gate-time will be $2T$ and  the total pulse area is $2\pi$. Nevertheless, there is still no specific restriction on the shapes of  $\Omega(t)$, $\Omega^I_1(t)$ and $\Omega^I_2(t)$,  except their integration. In this way, even in the presence of the $X$ error {during the whole evolution process, including in both the conventional NHQC process and the two inserted dynamical processes}, the dressed state $|b\rangle$ can still  go back to the original starting point after the cyclic evolution with high accuracy, as shown in Fig. \ref{fig1}(d), leading to robust  holonomic quantum gates.

We next proceed to analytically proof the above result. In the presence of  the \emph{X } error, {including in both the conventional NHQC process and the two inserted dynamical processes}, the error Hamiltonian will be $\mathcal{H}^\epsilon(t)$ as defined before, and the holonomic  gates in the dressed basis $\{|b\rangle, |d\rangle\}$ will change to
\begin{eqnarray}\label{Ue2}
U^\epsilon_c & = & \left [\cos^4\frac{\mu\pi}{2}+ \left (\sin^2\frac{\mu\pi}{2}
+\frac{1}{4}{\sin^2\mu\pi}\right)e^{i\gamma_g} \right ]|b\rangle \langle b|\notag\\
 &&  +|d\rangle \langle d|.
\end{eqnarray}
Then, when the \emph{X } error is small, i.e., $\epsilon \ll 1$, the gate-fidelity  in the dressed basis $\{|b\rangle, |d\rangle\}$ of the three-level quantum system will be
\begin{eqnarray}\label{Ue3}
F_c &=& \frac{|Tr(U^\dagger U^\epsilon_c)|}{|Tr(U^\dagger U)|}\notag\\
&=&\frac{1}{2}\left| 1+ \sin^2\frac{\mu\pi}{2} +\frac{1}{4}{\sin^2\mu\pi} +\cos^4\frac{\mu\pi}{2}e^{-i\gamma_g}  \right|     \notag\\
& \approx & 1-\epsilon^4\pi^4(1-\cos\gamma_g)/32,
\end{eqnarray}
which shows that  dynamically corrected holonomic  gates can improve the gate robustness against the  \emph{X} error from the second order in the conventional NHQC case, see Eq. (\ref{Ue1}), to the fourth order.  As shown in Fig. \ref{fig1}(d), in this case, the geometric trajectory  starting from the dressed state $ |b \rangle$, the evolution process is  driven by   $  \mathcal{H}^\epsilon(t) $ with \emph{X } errors, however, by inserting two dynamically corrected Hamiltonian, which also be with the \emph{X} error, the ruined evolution process is now be corrected. That is to say, the dressed state $ |b \rangle$ arriving at the dressed basis $ |a \rangle$ accurately after the first half of the evolution path, and then going accurately  back to the dressed basis $|b \rangle$ during the second  half, and  finally acquires a pure geometric phase.

\begin{figure}[tbp]
	\begin{center}
		\includegraphics[width=\columnwidth]{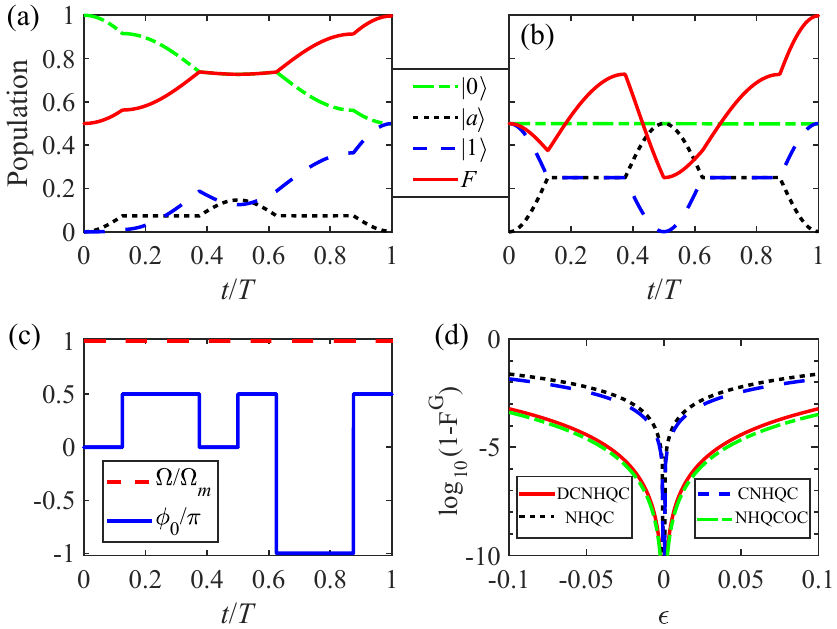}
\caption{The state population and fidelity dynamics for the   dynamically corrected holonomic (a)  \emph{H}   and (b) \emph{S}  gates. (c) The shapes of $\Omega$ and phases $\phi_0$ in $\mathcal{H}(t)$ for the \emph{S} gate. (d) The gate robustness with  logarithm to the basis 10 of the gate infidelity  for the \emph{S} gate under the \emph{X } error without decoherence, obtaining from DCNHQC, NHQCOC, conventional single-loop NHQC, and two-loop CNHQC, where the former twos show better performance.}\label{fig3}
	\end{center}
\end{figure}

Furthermore, to faithfully confirm our results, we consider a $V$ type three-level quantum system as shown in Fig. \ref{fig2}(a)  to simulate the gate performance of the proposed single-qubit gates using  the Lindblad master equation of
\begin{eqnarray}  \label{master}
\dot\rho_1 &=& i[\rho_1, \mathcal{H}]  + \frac{1}{2}  \sum^4_{n=1} \Gamma_n \mathcal{L}(\sigma_n),
\end{eqnarray}
where $\rho_1$  is the density matrix of the considered system and $\mathcal{L}(\sigma_n)=2\sigma_n\rho_1 \sigma_n^\dagger-\sigma_n^\dagger \sigma_n \rho_1 -\rho_1 \sigma_n^\dagger \sigma_n$ is the Lindbladian of the operator $\sigma_n$ with $\sigma_1=|a\rangle\langle 0|$, $\sigma_2=|a\rangle\langle 1|$, $\sigma_3=(|0\rangle\langle 0|-|a\rangle\langle a|)/ 2$,  $\sigma_4=(|1\rangle\langle 1|-|a\rangle\langle a|)/ 2$, and $\Gamma_n$ are their corresponding decoherence rates. For typical  systems that are suitable for physical implementation of quantum computation, the $\Gamma_n/\Omega_{m}$ ratio with $\Omega_m  = \text{ max} \{\Omega(t)\}$ can well below the  $10^{-4}$ value within current state-of-art technologies, even for the solid-state quantum systems which  typical have shorter coherent time. For example, for superconducting transmon qubits \cite{sqc, QS2019}, the qubit coherent time can approach 0.1 ms, corresponding to kHz level decoherence rate,  while the Rabi frequency can be tens of MHz. In our simulation, we set $\Omega(t)=\Omega_{m} = 1$, $\phi_0(0) = 0$, and the decoherence rates  $\Gamma_n=\Gamma= 5 \times 10^{-4} $. {Specifically, we numerically demonstrate the state population and fidelity dynamics of the dynamically corrected  Hadamard (\emph{H}) gate with $\theta = \pi/4, \gamma_g = \pi,  \phi=0$ and  \emph{S} gate with $\theta = 0, \gamma_g = \pi/2,  \phi=0$. The corresponding initial states are set to be $|\psi_i\rangle_H  = |0\rangle$  and  $|\psi_i\rangle_S  = (|0\rangle+|1\rangle)/\sqrt{2}$, respectively. And, we use the state fidelities, defined  by $F_{H/S} =_{H/S}\langle \psi_f|\rho_1|\psi_f\rangle_{H/S}$ with $|\psi_f\rangle_H = (|0\rangle+|1\rangle)/\sqrt{2}$ and  $|\psi_f\rangle_S = (|0\rangle+i|1\rangle)/\sqrt{2}$ being the corresponding ideal target states, to evaluate these two gates.  As shown in Figs. \ref{fig3}(a) and \ref{fig3}(b), we obtain the state fidelities for the \emph{H} and \emph{S} gate as $F_H=99.69\%$ and $F_S=99.73\%$, respectively. The amplitude envelopes  of $\Omega(t)$ and phase envelopes of $\phi_0$ of the Hamiltonian $\mathcal{H}(t)$ for \emph{S} gate are plotted  in Fig. \ref{fig3}(c). Besides, to fully evaluate the gate performance, using six initial states $|0\rangle, |1\rangle, (|0\rangle+|1\rangle)/\sqrt{2}, (|0\rangle-|1\rangle)/\sqrt{2}, (|0\rangle+i|1\rangle)/\sqrt{2}$ and $(|0\rangle-i|1\rangle)/\sqrt{2}$, we define gate fidelity as \cite{GateF}
\begin{equation}
F^G_1=\frac{1}{6}{\sum^{6}_{j = 1}} {_j\langle\psi_i|U^\dagger \rho_1 U|\psi_i\rangle_j}.
\end{equation}
The gate fidelities of the $\emph{H}$ and $\emph{S}$ gates are $F^G _H= 99.74\%$ and $F^G _S= 99.74\%$, respectively, and the infidelity is mainly due to the decoherence effect. Furthermore, we have numerically confirmed that  when $\Gamma \leq 2 \times 10^{-4}$, both gate fidelities can exceed 99.9\%.

 \begin{figure}[tbp]
	\begin{center}
\includegraphics[width=8cm]{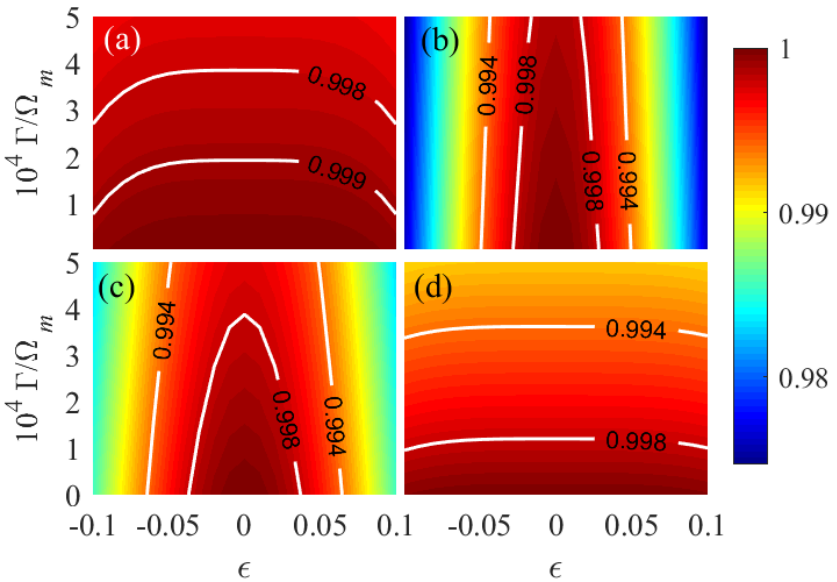}
\caption{Robustness  of the \emph{S}  gate  in  (a) DCNHQC, (b) conventional single-loop NHQC, (c) two-loop CNHQC, and (d) NHQCOC cases,  considering both the  \emph{X } error and the decoherence effect with uniformed decoherence rate $\Gamma/\Omega_{m}$. In this realistic case,  the proposed DCNHQC strategy performs best.}\label{fig4}
	\end{center}
\end{figure}

Finally, to demonstrate the robustness  of the dynamically corrected holonomic  gates, we consider the \emph{X } error within the range of $-0.1\leq\epsilon\leq0.1$, and select the \emph{S} gate for a demonstration purpose. Firstly, in the case of without decoherence, we numerically compared the robustness of the present DCNHQC, conventional single-loop NHQC \cite{SingleloopSQ}, two-loop CNHQC \cite{zhu2019}, and NHQCOC \cite{Li, ai2020} cases. As shown in Fig. \ref{fig3}(d), both  DCNHQC and NHQCOC can improve the gate robustness against the  \emph{X} error from the second order in the conventional NHQC case to the fourth order, which demonstrates  the stronger robustness of dynamically corrected holonomic  gates against  the \emph{X } error. Secondly, we repeat the above  comparison under  decoherence,   with uniform  rate of $\Gamma_n=\Gamma \in [0,5] \times 10^{-4}$. As shown in Fig. \ref{fig4}, after considering the influence of both  the \emph{X } error and the decoherence effect, the current DCNHQC protocol performs best. In this case, the gate fidelity in the NHQCOC case will be decreased rapidly as it suffers severely from the decoherence effect, due to its longer gate-time required by deliberate engineered pulse shapes.   In addition, from Fig. \ref{fig4}(a), when $\Gamma \leq  0.9\times 10^{-4}$, the fidelity for all gate can be higher than $99.9\%$ for the \emph{X } error within the range of $-0.1\leq\epsilon\leq0.1$, thus, our proposal is apparently more robust than previous ones.

\section{Physical implementation}
In the last section, we show that, based on the dynamically corrected technique, the \emph{X } error can be well suppressed for dynamically corrected holonomic gates. Meanwhile,  when incorporating  a DFS encoding for the collective dephasing noise, our scheme can also be robust against the collective \emph{Z} error. We now turn to consider the physical implementation of the scheme combining the dynamically correction technique and the DFS encoding. We consider three resonantly coupled two-level qubits with conventional two-body exchange interaction as shown in Fig. \ref{fig2}(b), the Hamiltonian of which is
\begin{eqnarray}\label{H1}
  \mathcal{H}_1(t) &=& J_1(t)e^{-\text{i}\varphi_1} S^+_1S ^-_{A_1}+J_2(t)e^{-\text{i}\varphi_2} S^+_2S ^-_{A_1}+\textrm{H.c.},\notag\\
\end{eqnarray}
where $J_{k}(t) $ $(k=1,2)$ are  coupling strengths between qubits $Q_k$ and $Q_{A_1}$, and $S^+ = |1\rangle\langle0|$ is the creation operator. This exchange interaction Hamiltonian is very general and can be implemented in trapped ions \cite{liang2014}, cavity coupled NV centers \cite{zhouj}, different superconducting quantum circuit setups \cite{xue1, xue2, wangym2016, xue3, xue4, Chentoc3}, etc. Then,  we combine three physical qubits and encode them into a logical qubit that belongs to a three-dimensional DFS $\mathcal{S}_1= $ Span$\{|100\rangle, |010\rangle, |001\rangle\}$ with $|100\rangle\equiv |1\rangle_1\otimes|0\rangle_2\otimes|0\rangle_{A_1}$. In this single-excitation subspace, logical qubit states are defined as $\{|0\rangle_L = |100\rangle$ and $|1\rangle_L = |010\rangle\}$,  with $|A_1\rangle_L = |001\rangle$ serving as an ancillary state. Assuming we want to encode an   arbitrary state on physical qubit 1, e.g., $|\psi\rangle_1=a|1\rangle_{1}+b|0\rangle_{1}$, into the logical qubit subspace, i.e., $|\psi\rangle_L=a|100\rangle +b|010\rangle=a|0\rangle_{L}+b|1\rangle_{L}$. When  the other two qubits are both in their ground states, this can be achieved as following. First, a NOT gate is applied on the qubit 2, and then a CNOT gate is applied on qubits 1 and 2, with qubit 1 being the control qubit. Meanwhile, the decoding circuit can be obtained by inversing these two steps.  In the  logical basis, 
Eq. (\ref{H1}) can be rewritten as
 \begin{equation}\label{H1L}
  \mathcal{H}^1_L(t) = J_1(t) e^{-\text{i}\varphi_1} |0\rangle_L\langle A_1|+J_2(t) e^{-\text{i}\varphi_2} |1\rangle_L\langle A_1|+\textrm{H.c.},
\end{equation}
which has the same form as the Hamiltonian $\mathcal{H}(t)$ in Eq. (\ref{HI}), and thus can naturally be used to apply the DCNHQC strategy to suppress the $X$-error.  When $\Gamma = 2 \times 10^{-4}$, the fidelity of the $H$ gate is 99.85\%, where the infidelity is mainly due to the decoherence effect of physical qubits.

 \begin{figure}[tbp]
	\begin{center}
		\includegraphics[width=8cm]{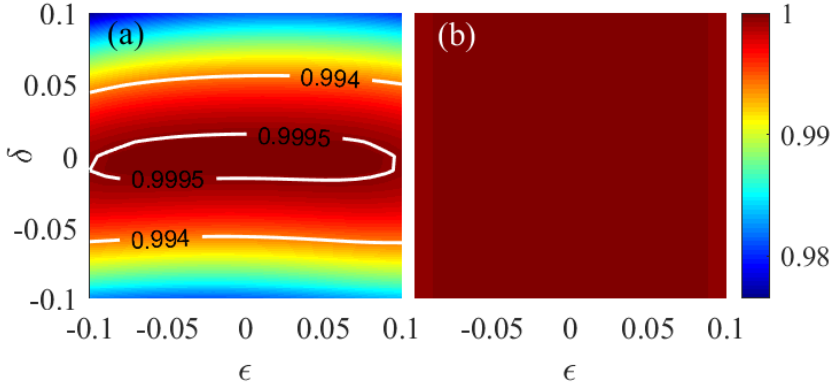}
\caption{Gate  robustness  of the \emph{S}  gate  in DCNHQC (a) without and (b)  with the DFS encoding,  considering both the  \emph{X } error and the  collective \emph{Z } error, where the encoded case performs much better.}\label{fig6}
	\end{center}
\end{figure}

Here, we also compare  the gate robustness of our  DCNHQC without and  with the DFS encoding in Fig. \ref{fig6}, considering both the  \emph{X } and  \emph{Z } errors for $\mathcal{H}^E(t) =  (1+\epsilon)\mathcal{H}(t) - \delta \Omega_m|a\rangle\langle a|$
and {$\mathcal{H}^E_L(t) =  (1+\epsilon)\mathcal{H}^1_L(t)- \delta\Omega_m (|0\rangle_L\langle0|+|1\rangle_L\langle1|+|A_1\rangle_L\langle A_1|)$,}
respectively. We consider the \emph{S} gate for a typical example.  As shown in Fig. \ref{fig6}(a),  the gate from DCNHQC can only suppress the $X$ error. While combing it with the DFS encoding, the gate with encoding can suppress both the  \emph{X }  and   \emph{Z } errors, and the gate infidelities are well within $0.1\%$ in the considered error range, as shown in Fig. \ref{fig6}(b). Here, for the numerical simulation of the DFS encoding case, we consider a simple case that  the dephasing is collective, i.e., the \emph{Z } error is the same for each physical qubit. When the \emph{Z } errors $\delta_j$ are different in physical qubits, the DFS encoding can still eliminate the overlapping part of the $\delta_j$  drifts, which are of the low-frequency nature for solid-state qubits and thus can be regarded as constants during a certain quantum gate operation.

In addition, to realize universal quantum computation, a nontrivial  two-qubit gate is also necessary. Next, we consider six two-level physical qubit systems with effectively and resonantly coupling, see Fig. \ref{fig2}(b), to implement nontrivial  two-qubit gates. The considered interaction Hamiltonian is
\small
\begin{eqnarray}\label{H2}
  \mathcal{H}_2(t) &=& g_1(t) e^{-\text{i}\varphi_3} S^+_2S ^-_3+g_2(t) e^{-\text{i}\varphi_4} S^+_2S ^-_4+\textrm{H.c.}, 
\end{eqnarray}
where $g_{k}(t) $ $(k=1,2)$ are  coupling strengths between qubit pairs $Q_2$, $Q_3$ and $Q_2$, $Q_4$ respectively. Stick to the single-qubit DFS encoding, there exists a six-dimensional DFS $\mathcal{S}_2= $ Span$\{|100100\rangle$, $|100010\rangle$, $|010100\rangle$, $|010010\rangle$, $|110000\rangle$,  $|000110\rangle\}$ with $|100100\rangle\equiv |1\rangle_1 \otimes|0\rangle_2 \otimes|0\rangle_{A_1} \otimes|1\rangle_3 \otimes|0\rangle_4 \otimes|0\rangle_{A_2}$, and logical qubits are defined as $\{|00\rangle_L = |100100\rangle, |01\rangle_L = |100010\rangle, |10\rangle_L = |010100\rangle, |11\rangle_L = |010010\rangle\}$ with $|A_1\rangle_L = |110000\rangle$ and $|A_2\rangle_L = |000110\rangle$ being two  auxiliary states. In this  two-qubit DFS subspace, the Hamiltonian of the coupled system in Eq. (\ref{H2}) changes to
 \begin{eqnarray}\label{H2L}
  \mathcal{H}^2_L(t) &=& g_1(t) e^{-\text{i}\varphi_3} | A_1 \rangle_L\langle 00|+g_2(t) e^{-\text{i}\varphi_4} | A_1 \rangle_L\langle 01|\notag\\
& +& g_1(t)e^{-\text{i}\varphi_3}  | 11\rangle_L\langle A_2|+g_2(t)e^{-\text{i}\varphi_4} | 10 \rangle_L\langle A_2| \notag \\
& +& \textrm{H.c.},
\end{eqnarray}
which has two three-level structure, in subspaces Span$\{|00\rangle_L, |01\rangle_L, |A_1\rangle_L\}$ and Span$\{|10\rangle_L, |11\rangle_L, |A_2\rangle_L\}$ \cite{NJP}, each of which has the same form with the Hamiltonian $\mathcal{H}(t)$ in Eq. (\ref{HI}). Then, we can also implement dynamically corrected nonadiabatic  holonomic two-qubit gates, which suppresses  both the \emph{X} and  \emph{Z} errors.

 \begin{figure}[tbp]
	\begin{center}
\includegraphics[width=7.5cm]{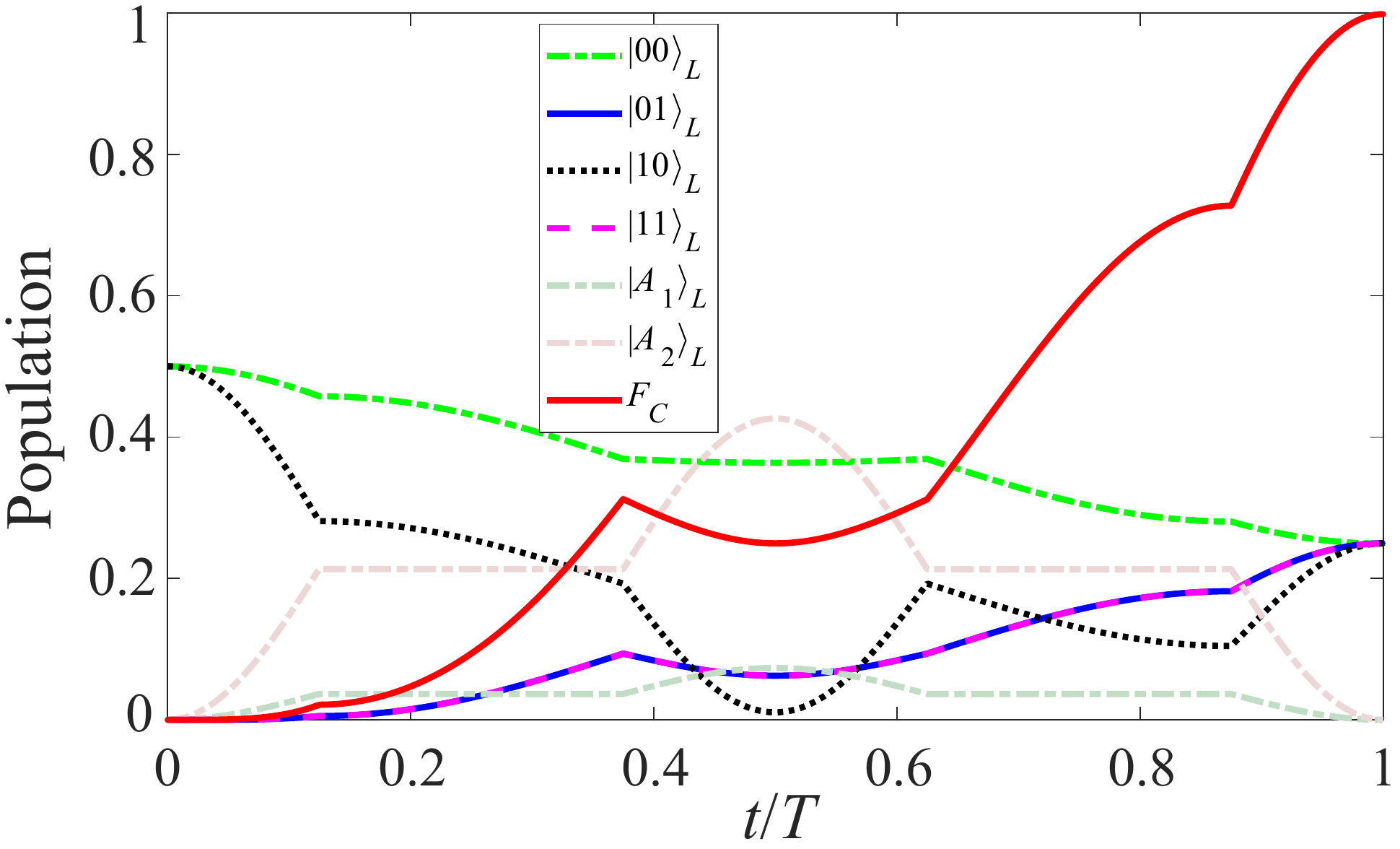}
\caption{State dynamics of gate $C$ for an initial state of $(|0\rangle_L+|1\rangle_L)|0\rangle_L/\sqrt{2}$.}\label{fig7}
	\end{center}
\end{figure}

Specifically, besides the two inserted Hamiltonian, when $\int_0 ^\tau \sqrt{g_1(t)^2 +g_2(t)^2} dt=\pi$ and geometric phase is $\pi$, the evolution operator in subspaces Span$\{|00\rangle_L, |01\rangle_L, |A_1\rangle_L,|10\rangle_L, |11\rangle_L, |A_2\rangle_L\}$ for Hamiltonian in Eq. (\ref{H2L}) can be written as
\begin{eqnarray} \label{two-qubit1}
&&U_T(\eta, \varphi)  \notag\\
&&=\left(
  \begin{array}{cccccc}
    \cos\eta & \sin\eta e^{-i\varphi} & 0 & 0 & 0 & 0  \\
    \sin\eta e^{i\varphi} & -\cos\eta & 0 & 0 & 0 & 0  \\
    0 & 0 & -i &  0 & 0 & 0  \\
    0 & 0 & 0 & -\cos\eta & \sin\eta e^{-i\varphi} & 0 \\
    0 & 0 & 0 & \sin\eta e^{i\varphi} & \cos\eta & 0 \\
    0 & 0 & 0 &  0 & 0 & -i  \\
  \end{array}
\right),\notag\\
\end{eqnarray}
where $\tan (\eta/2)= g_1(t)/ g_2(t)$ with $\eta$ being a constant angle and $\varphi=\varphi_3 - \varphi_4 +\pi$. Notably, there is no transitions between the two-logical-qubit subspace Span$\{|00\rangle_L, |01\rangle_L,|10\rangle_L, |11\rangle_L\}$ and the auxiliary subspace Span$\{ |A_1\rangle_L, |A_2\rangle_L\}$, thus, the evolution operator
in the two logical-qubit subspace    reduces to  \cite{NJP}
\begin{eqnarray} \label{two-qubit2}
U_2(\eta, \varphi)=\left(
  \begin{array}{cccc}
    \cos\eta & \sin\eta e^{-i\varphi} & 0 & 0 \\
    \sin\eta e^{i\varphi} & -\cos\eta & 0 & 0 \\
    0 & 0 & -\cos\eta & \sin\eta e^{-i\varphi} \\
    0 & 0 & \sin\eta e^{i\varphi} & \cos\eta \\
  \end{array}
\right).\notag\\
\end{eqnarray}

Notably, the operations in two subspaces of $\{|00\rangle_L , |01\rangle_L\}$ and $\{|10\rangle_L, |11\rangle_L \}$  are different, thus the operator in Eq. (\ref{two-qubit2}) denotes a nontrivial two-qubit gate in general. For example, $U_2(\pi/4, 0)\equiv C$ is an entangling gate \cite{xue2}, and
\begin{eqnarray}
C\times(I_1\otimes H_2)&=&\left(
  \begin{array}{cccc}
      1 & 0 & 0 & 0 \\
0 & 1 & 0 & 0 \\
    0 & 0 & 0 & -1 \\
    0 & 0 &1 & 0\\
    \end{array}
\right)\notag\\
&=&\text{CNOT} \times \text{CP},
\end{eqnarray}
with CNOT and CP being the controlled not and controlled phase gates, respectively. Furthermore, when $\Gamma = 2 \times 10^{-4}$, as shown in Fig. \ref{fig7}, we simulate state dynamics of gate $C$ governed by Eq. (\ref{H2}) with an initial product  state of $(|0\rangle_L+|1\rangle_L)|0\rangle_L/\sqrt{2}$ and the state fidelity being $F_C = 99.82\% $. In this case, the corresponding final state will be $$(|00\rangle_L+|01\rangle_L-|10\rangle_L+|11\rangle_L)/2=(|-0\rangle_L+|+1\rangle_L)/\sqrt{2}$$ with $|\pm\rangle=(|0\rangle_L\pm|1\rangle_L)/\sqrt{2}$, which is an entangled state. As our manipulation of the logical-qubit subspace only involves the single-excitation subspace of governed Hamiltonian in Eqs. (\ref{H1}) and (\ref{H2}), the physical qubit leakage and other high-order oscillating terms can be effectively suppressed, which leads to negligible gate-infidelity. Meanwhile, due to the DFS encoding,  the dephasing induced gate-infidelity is also negligible, less than 0.01\%. Thus, the gate-infidelity here is  mainly due to the  qubit decay effect, about 0.18\%, which can be further suppressed by improving the lifetime of the physical qubit or shortening the gate-time.

\section{Conclusion}
In conclusion,  we propose a general scheme for universal NHQC  with the dynamical correction technique,  which can improve gate robustness against the  \emph{X} error from second order of conventional NHQC case to the fourth order,  and thus retaining the main merit of geometric quantum gates. Numerical simulation shows that our current  protocol can be  better than previous implementations.   Remarkably, when incorporating  a DFS encoding, our scheme can also be robust against  the \emph{Z} errors. Finally, we present the physical implementation of the dynamically corrected protocol with DFS encoding using exchange interactions, which is very conventional in solid-state quantum systems. Therefore, our scheme provides a promising strategy for the future scalable  fault-tolerant quantum computation.


\acknowledgements
This work was supported by the Key-Area Research and Development Program of GuangDong Province (Grant No. 2018B030326001), the National Natural Science Foundation of China (Grant No. 11874156), and the Science and Technology Program of Guangzhou (No. 2019050001).


\begin{thebibliography}{99}
\bibitem{shor} P. W. Shor, in \emph{Proceedings of the 35th Annual Symposium on the Foundations of Computer Science} (IEEE Press, Los Alamitos, CA, 1994).

%


\bibitem{QS2019}
F. Arute, K. Arya, R. Babbush, D. Bacon, J. C. Bardin, R. Barends, R. Biswas, S. Boixo, F. G. S. L. Brandao, D. A. Buell \emph{et al}.,
Quantum supremacy using a programmable superconducting processor,
Nature (London) \textbf{574}, 505 (2019).

\bibitem{Abelian}
M. V. Berry,
Quantal phase factors accompanying adiabatic changes,
Proc. R. Soc. London A \textbf{392}, 45 (1984).

\bibitem{non-Abelian}
F. Wilczek and A. Zee,
Appearance of gauge structure in simple dynamical systems,
Phys. Rev. Lett. \textbf{52}, 2111 (1984).

\bibitem{AA}
Y. Aharonov and J. Anandan,
Phase change during a cyclic quantum evolution,
Phys. Rev. Lett. \textbf{58}, 1593 (1987).

\bibitem{zanardi}
P. Zanardi and M. Rasetti,
Holonomic quantum computation,
Phys. Lett. A \textbf{264}, 94 (1999).

\bibitem{AGQC1}
J. Pachos, P. Zanardi, and M. Rasetti,
Non-Abelian Berry connections for quantum computation,
Phys. Rev. A \textbf{61}, 010305(R) (1999).

\bibitem{Duan}
L.-M. Duan, J. I. Cirac, and P. Zoller,
Geometric manipulation of trapped ions for quantum computation,
Science \textbf{292}, 1695 (2001).


\bibitem{cenlx} L.X. Cen, X.Q. Li, Y.J. Yan, H.Z. Zheng, and S.J. Wang,
Evaluation of Holonomic Quantum Computation: Adiabatic Versus Nonadiabatic,
Phys. Rev. Lett. {\bf 90}, 147902 (2003).

\bibitem{NJP}
E. Sj\"{o}qvist, D. M. Tong, L. M. Andersson, B. Hessmo, M. Johansson, and K. Singh,
Non-adiabatic holonomic quantum computation,
New J. Phys. \textbf{14}, 103035 (2012).

\bibitem{TongDM}
G. F. Xu, J. Zhang, D. M. Tong, E. Sj\"{o}qvist, and L. C. Kwek,
Nonadiabatic holonomic quantum computation in decoherence-free subspaces,
Phys. Rev. Lett. \textbf{109}, 170501 (2012).



\bibitem{surface1} Y.-C. Zheng and T. A. Brun,
Fault-tolerant holonomic quantum computation in surface codes,
Phys. Rev. A {\bf 91}, 022302 (2015).

\bibitem{Singleloopxu} G. F. Xu, C. L. Liu, P. Z. Zhao, and D. M. Tong,
Nonadiabatic holonomic gates realized by a single-shot implementation,
Phys. Rev. A {\bf 92}, 052302 (2015).

\bibitem{Singleloop}
E.~Herterich and E.~Sj\"{o}qvist,
Single-loop multiple-pulse nonadiabatic holonomic quantum gates,
Phys. Rev. A \textbf{94}, 052310 (2016).


\bibitem{SingleloopSQ}
Z.-P. Hong, B.-J. Liu, J.-Q. Cai, X.-D. Zhang, Y. Hu, Z. D. Wang, and Z.-Y. Xue,
Implementing universal nonadiabatic holonomic quantum gates with transmons,
Phys. Rev. A \textbf{97}, 022332 (2018).



\bibitem{liubj}
B.-J. Liu, Z.-H. Huang, Z.-Y. Xue, and X.-D. Zhang,
Superadiabatic holonomic quantum computation in cavity QED,
Phys. Rev. A 95, 062308 (2017).


\bibitem{surface2} J. Zhang, S. J. Devitt, J. Q. You, and F. Nori,
Holonomic surface codes for fault-tolerant quantum computation,
Phys. Rev. A 97, 022335 (2018).

\bibitem{eric} N. Ramberg and E. Sj\"{o}qvist,
Environment-Assisted Holonomic Quantum Maps,
Phys. Rev. Lett. {\bf 122}, 140501 (2019).

\bibitem{surface3} C. Wu, Y. Wang, X.-L. Feng, and J.-L. Chen,
Holonomic Quantum Computation in Surface Codes,
Phys. Rev. Appl. {\bf 13}, 014055 (2020).

\bibitem{zhaopz2020} P. Z. Zhao, K. Z. Li, G. F. Xu, and D. M. Tong,
General approach for constructing Hamiltonians for nonadiabatic holonomic quantum computation,
Phys. Rev. A  {\bf 101}, 062306 (2020).

\bibitem{liubj2020} B.-J. Liu and M.-H. Yung,
Leakage Suppression for Holonomic Quantum Gates,
Phys. Rev. Appl. {\bf 14}, 034003 (2020).


\bibitem{Abdumalikov35}
A.~A. Abdumalikov, J.~M. Fink, K.~Juliusson, M.~Pechal, S.~Berger, A.~Wallraff, and S.~Filipp,
Experimental realization of non-Abelian non-adiabatic geometric gates,
Nature (London) \textbf{496}, 482 (2013). 

\bibitem{Feng39}
G.~Feng, G.~Xu, and G.~Long,
Experimental realization of nonadiabatic holonomic quantum computation,
Phys. Rev. Lett. \textbf{110}, 190501 (2013).



\bibitem{Zu41}
C.~Zu, W.-B. Wang, L.~He, W.-G. Zhang, C.-Y. Dai, F.~Wang, and L.-M. Duan,
Experimental realization of universal geometric quantum gates with solid-state spins,
Nature (London) \textbf{514}, 72 (2014).   

\bibitem{AC2014}
S. Arroyo-Camejo, A. Lazariev, S. W. Hell, and G. Balasubramanian,
Room temperature high-fidelity holonomic single-qubit gate on a solid-state spin,
Nat. Commun. \textbf{5}, 4870 (2014).

\bibitem{nv2017}
Y. Sekiguchi, N. Niikura, R. Kuroiwa, H. Kano, and H. Kosaka,
Optical holonomic single quantum gates with a geometric spin under a zero field,
Nat. Photonics \textbf{11}, 309 (2017).

\bibitem{nv20172}
B. B. Zhou, P. C. Jerger, V. O. Shkolnikov, F. J. Heremans, G. Burkard, and D. D. Awschalom,
Holonomic quantum control by coherent optical excitation in diamond,
Phys. Rev. Lett. \textbf{119}, 140503 (2017).

\bibitem{li2017}
H. Li, L. Yang, and G. Long,
Experimental realization of single-shot nonadiabatic holonomic gates in nuclear spins,
Sci. China: Phys., Mech. Astron. \textbf{60}, 080311(2017).

\bibitem{xuy37}
Y. Xu, W. Cai, Y. Ma, X. Mu, L. Hu, T. Chen, H. Wang, Y. P. Song, Z.-Y. Xue, Z.-Q. Yin, and L. Sun,
Single-Loop Realization of Arbitrary Nonadiabatic Holonomic Single-Qubit Quantum Gates in a Superconducting Circuit,
Phys. Rev. Lett. \textbf{121}, 110501 (2018).


\bibitem{three1}  S. B. Zheng, C. P. Yang, and F. Nori,
Comparison of the sensitivity to systematic errors between nonadiabatic non-Abelian geometric gates and their dynamical counterparts,
Phys. Rev. A 93, 032313 (2016).

\bibitem{three2}  J. Jing, C.-H. Lam, and L.-A. Wu,
Non-Abelian holonomic transformation in the presence of classical noise,
Phys. Rev. A 95, 012334 (2017).


\bibitem{encode}  Y.-C. Zheng and T. A. Brun,
Fault-tolerant scheme of holonomic quantum computation on stabilizer codes with robustness to low-weight thermal noise,
Phys. Rev. A {\bf 89}, 032317 (2014).


\bibitem{zhangj2014}
J. Zhang, L.-C. Kwek, E. Sj\"{o}qvist, D. M. Tong, and P. Zanardi,
Quantum computation in noiseless subsystems with fast non-Abelian holonomies,
Phys. Rev. A {\bf 89}, 042302 (2014).

\bibitem{liang2014}  Z.-T. Liang, Y.-X. Du, W. Huang, Z.-Y. Xue, and H. Yan,
Nonadiabatic holonomic quantum computation in decoherence-free subspaces with trapped ions,
Phys. Rev. A {\bf 89}, 062312 (2014).


\bibitem{zhouj} J. Zhou, W.-C. Yu, Y.-M. Gao, and Z.-Y. Xue,
Cavity QED implementation of non-adiabatic holonomies for universal quantum gates in decoherence-free subspaces with nitrogen-vacancy centers,
Opt. Express {\bf 23}, 14027 (2015).

\bibitem{xue1} Z.-Y. Xue, J. Zhou, and Z. D. Wang,
Universal holonomic quantum gates in decoherence-free subspace on superconducting circuits,
Phys. Rev. A {\bf 92}, 022320 (2015).

\bibitem{wangym2016} Y. Wang, J. Zhang, C. Wu, J. Q. You, and G. Romero,
Holonomic quantum computation in the ultrastrong-coupling regime of circuit QED,
Phys. Rev. A {\bf 94}, 012328 (2016).

\bibitem{xue2} Z.-Y. Xue, J. Zhou, Y.-M. Chu, and Y. Hu,
Nonadiabatic holonomic quantum computation with all-resonant control,
Phys. Rev. A {\bf 94}, 022331 (2016).


\bibitem{xue3} Z.-Y. Xue, F.-L. Gu, Z.-P. Hong, Z.-H. Yang, D.-W. Zhang, Y. Hu, and J. Q. You,
Nonadiabatic Holonomic Quantum Computation with Dressed-State Qubits,
Phys. Rev. Appl. {\bf 7}, 054022 (2017).

\bibitem{zhaopz2017} P. Z. Zhao, G. F. Xu, Q. M. Ding, E. Sj\"{o}qvist, and D. M. Tong,
Single-shot realization of nonadiabatic holonomic quantum gates in decoherence-free subspaces,
Phys. Rev. A {\bf 95}, 062310 (2017).

\bibitem{xue4} L.-N. Ji, T. Chen, and Z.-Y. Xue,
Scalable nonadiabatic holonomic quantum computation on a superconducting qubit lattice,
Phys. Rev. A {\bf 100}, 062312 (2019).


\bibitem{wangym2020} Y. Wang, Y. Su, X. Chen, and C. Wu,
Dephasing-Protected Scalable Holonomic Quantum Computation on a Rabi Lattice,
Phys. Rev. Appl. {\bf 14}, 044043 (2020).


\bibitem{composite} G. F. Xu, P. Z. Zhao, T. H. Xing, E. Sj\"{o}qvist, and D. M. Tong,
Composite nonadiabatic holonomic quantum computation,
Phys. Rev. A 95, 032311 (2017).

\bibitem{zhu2019}
Z. Zhu, T. Chen, X. Yang, J. Bian, Z.-Y. Xue, and X. Peng,
Single-loop and composite-loop realization of nonadiabatic holonomic quantum gates in a decoherence-free subspace,
Phys. Rev. Appl. \textbf{12}, 024024 (2019).

\bibitem{dd} Y. Sekiguchi, Y. Komura, and H. Kosaka,
Dynamical Decoupling of a Geometric Qubit,
Phys. Rev. Appl. {\bf 12}, 051001 (2019).

\bibitem{dd2}  X. Wu and P. Z. Zhao,
Universal nonadiabatic geometric gates protected by dynamical decoupling,
Phys. Rev. A {\bf 102}, 032627 (2020).


\bibitem{Liu18}
B.-J. Liu, X.-K. Song, Z.-Y. Xue, X. Wang, and M.-H.Yung,
Plug-and-Play Approach to Nonadiabatic Geometric Quantum Gates,
Phys. Rev. Lett. \textbf{123}, 100501 (2019).


\bibitem{yan2019}
T. Yan, B.-J. Liu, K. Xu, C. Song, S. Liu, Z. Zhang, H. Deng, Z. Yan, H. Rong, K. Huang, M.-H. Yung, Y. Chen, and D. Yu,
Experimental realization of nonadiabatic shortcut to non-Abelian geometric gates,
Phys. Rev. Lett. \textbf{122}, 080501 (2019).

\bibitem{Li}
S. Li, T. Chen, and Z.-Y. Xue,
Fast holonomic quantum computation on superconducting circuits with optimal control,
Adv. Quantum Technol. \textbf{3}, 2000001 (2020).



\bibitem{ai2020} M.-Z. Ai, S. Li, Z. Hou, R. He, Z.H.Qian, Z.-Y. Xue, J.-M. Cui, Y.-F. Huang, C.-F. Li, and G.-C. Guo,
Experimental realization of nonadiabatic holonomic single-qubit quantum gates with optimal control in a trapped ion,
Phys. Rev. Appl. {\bf 14}, 054062 (2020).

\bibitem{ai2021} M.-Z. Ai \emph{et al}., R. He, S. Li,  Z.-Y. Xue, J.-M. Cui, Y.-F. Huang, C.-F. Li, and G.-C. Guo,
Experimental Realization of Nonadiabatic Holonomic Single-Qubit Quantum Gates with Two Dark Paths in a Trapped Ion,
arXiv:2101.07483.



\bibitem{GPC}  B.-J. Liu, Y.-S. Wang, and M.-H. Yung,
Global Property Condition-based Non-adiabatic Geometric Quantum Control,
arXiv:2008.02176.


\bibitem{xugf2018} G. F. Xu, D. M. Tong, and E. Sj\"{o}qvist,
Path-shortening realizations of nonadiabatic holonomic gates,
Phys. Rev. A {\bf 98}, 052315 (2018).

\bibitem{zhang2019} F. Zhang, J. Zhang, P. Gao, and G. Long,
Searching nonadiabatic holonomic quantum gates via an optimization algorithm,
Phys. Rev. A {\bf 100}, 012329 (2019).


\bibitem{Chentoc3} T. Chen, P. Shen, and Z.-Y. Xue,
Robust and Fast Holonomic Quantum Gates with Encoding on Superconducting Circuits,
Phys. Rev. Appl. {\bf 14}, 034038 (2020).

\bibitem{BNHQC}
B.-J. Liu, Z.-Y. Xue, and M.-H.Yung,
Brachistochronic Non-Adiabatic Holonomic Quantum Control,
arXiv:2001.05182.

\bibitem{yuyang} Z. Han, Y. Dong, B. Liu, X. Yang, S. Song, L. Qiu, D. Li, J. Chu, W. Zheng, J. Xu, T. Huang, Z. Wang, X. Yu, X. Tan, D. Lan, M.-H. Yung, and  Y. Yu,
Experimental Realization of Universal Time-optimal non-Abelian Geometric Gates,
arXiv:2004.10364.


\bibitem{dc1} K. Khodjasteh and L. Viola,
Dynamically Error-Corrected Gates for Universal Quantum Computation,
Phys. Rev. Lett. \textbf{102}, 080501 (2009).


\bibitem{dc2}  X. Wang, L. S. Bishop, J. P. Kestner, E. Barnes, K. Sun, and S. D. Sarma,
Composite pulses for robust universal control  of singlet-triplet qubits,
Nat. Commun. \textbf{3}, 997 (2012).


\bibitem{dc3}  X. Rong, J. Geng, F. Shi, Y. Liu, K. Xu, W. Ma, F. Kong, Z. Jiang, Y. Wu, and J. Du,
Experimental fault-tolerant universal quantum gates with solid-state spins under ambient conditions,
Nat. Commun. \textbf{6}, 8748 (2015).


\bibitem{dfs1}
L.-M. Duan and G.-C. Guo,
Preserving coherence in quantum computation by pairing quantum bits,
Phys. Rev. Lett. \textbf{79}, 1953 (1997).

\bibitem{dfs2}
P. Zanardi and M. Rasetti,
Noiseless quantum codes,
Phys. Rev. Lett. \textbf{79}, 3306 (1997).

\bibitem{dfs3}
D. A. Lidar, I. L. Chuang, and K. B. Whaley,
Decoherence-free subspaces for quantum computation,
Phys. Rev. Lett. \textbf{81}, 2594 (1998).


\bibitem{shen} P. Shen, T. Chen, and Z.-Y. Xue, Ultrafast holonomic quantum gates, Phys. Rev. Appl. {\bf 16}, 044004 (2021).


\bibitem{sqc}
M. H. Devoret and R. J. Schoelkopf,
Superconducting circuits for quantum information: An outlook,
Science \textbf{339}, 1169 (2013).

\bibitem{GateF}
A. Klappenecker and M. Rotteler,
Mutually unbiased bases are complex projective 2-designs,
Proc. Int. Symp. Inf. Theory \textbf{2005}, 1740 (2005).

%

%
%
%
%

\end{thebibliography}
\end{document}